# Reinforcing increase of $\Delta T_c$ in MgB$_2$ smart meta-superconductors by adjusting the concentration of inhomogeneous phases

Yongbo Li, Guangyu Han, Hongyan Zou, Li Tang, Honggang Chen and Xiaopeng Zhao*

Smart Materials Laboratory, Department of Applied Physics, Northwestern Polytechnical University, Xi'an 710072, China;

*Correspondence: xpzhao@nwpu.edu.cn (X.Z.)

**Abstract**

Incorporating with inhomogeneous phases with high electroluminescence (EL) intensity to prepare smart meta-superconductors (SMSCs) is an effective method of increasing the superconducting transition temperature ($T_c$) and has been confirmed in both MgB$_2$ and Bi(Pb)SrCaCuO systems. However, the increase of $\Delta T_c$ ($\Delta T_c = T_c - T_{cpure}$) has been quite small because of the low optimal concentrations of inhomogeneous phases. In this work, three kinds of MgB$_2$ raw materials, namely, $^a$MgB$_2$, $^b$MgB$_2$, and $^c$MgB$_2$, were prepared with particle sizes decreasing in order. Inhomogeneous phases, Y$_2$O$_3$:Eu$^{3+}$ and Y$_2$O$_3$:Eu$^{3+}$/Ag, were also prepared and doped into MgB$_2$ to study the influence of doping concentration on the $\Delta T_c$ of MgB$_2$ with different particle sizes. Results show that reducing the MgB$_2$ particle size increases the optimal doping concentration of inhomogeneous phases, thereby increasing $\Delta T_c$. The optimal doping concentrations for $^a$MgB$_2$, $^b$MgB$_2$, and $^c$MgB$_2$ are 0.5%, 0.8%, and 1.2%, respectively. The corresponding $\Delta T_c$ values are 0.4, 0.9, and 1.2 K, respectively. This work open a new approach to reinforcing increase of $\Delta T_c$ in MgB$_2$ SMSCs.

## 1 Introduction

According to BCS theory, McMillan theoretically calculated the upper limit of the critical temperature ($T_c$) of conventional BCS superconductors to be 40 K, which is called the McMillan limit temperature [1,2]. Although the $T_c$ of conventional superconductors has an upper limit, the search for high-$T_c$ superconducting materials has been continuous. High-temperature superconductors [3–6], iron-based superconductors [7–10], high-pressure superconductors [11–15], and photo-induced superconductors [16,17] have been gradually studied and discovered. However, these superconducting materials are not simple conventional superconductors. Breaking the McMillan limit temperature remains a challenge for conventional BCS superconductors. In 2001, the superconductivity of MgB$_2$ was discovered [18]. The excellent superconductivity, simple preparation process, and especially high $T_c$ of $MgB_2$ quickly aroused great interest in the scientific community and led scholars to believe that the McMillan limit temperature may finally be surpassed [19–24]. Various methods have been applied to improve the superconductivity of $MgB_2$, which would not only improve the practical application of $MgB_2$ but also help transcend the McMillan limit temperature and further elucidate the superconducting mechanism [25–29]. Chemical doping is often used to study superconductivity. Unfortunately, many experimental results [30–35] confirm that this method reduces the $T_c$ of $MgB_2$. Thus far, no useful strategy for improving the $T_c$ of $MgB_2$ is yet available.

Metamaterial mainly refers to materials made up of two or more media, which can produce new properties that are not found in a single medium. Meta-method is often used to achieve some special properties and provides new ways of improving the $T_c$ of materials [36–38]. In 2007, our group proposed a method based on the structural design of metamaterials for increasing the $T_c$ of superconductors [39,40]. In this method, electroluminescence (EL) materials are directly doped into a superconductor to form a smart meta-superconductor (SMSC). The external field added during the measurement of the $T_c$ of SMSC with a four-probe method can excite the inhomogeneous phases to generate EL, achieving the purpose of strengthening the Cooper pairs, resulting the change of $T_c$ in macroscopic. A SMSC is a material whose $T_c$ can be adjusted and improved by the stimulus of external field, which is a new property and cannot be achieved by traditional doping with a second phase [41–46]. Our group subsequently conducted a series of studies, mainly using $MgB_2$ as the base superconducting material and $Y_2O_3:Eu^{3+}$ as the base EL material [41–43]. The results obtained in these studies show that unlike conventional chemical doping, which consistently reduces the $T_c$ of $MgB_2$, the SMSC method of doping EL materials could help increase the $T_c$ of $MgB_2$. The same conclusions were drawn from substituting the inhomogeneous phase with $YVO_4:Eu^{3+}$ or luminescent nanocomposite $Y_2O_3:Eu^{3+}/Ag$ [44,45] and replacing $MgB_2$ with Bi(Pb)SrCaCuO [46]. The effectiveness of improving the $T_c$ of superconducting materials through the SMSC method by doping with EL inhomogeneous phases has been proven, but the $\Delta T_c$ ($\Delta T_c = T_c - T_{cpure}$) values obtained are generally small (0.2–0.4 K). Our previous results show that the SMSC method can only improve $T_c$ at low concentrations of inhomogeneous phases and leads to a small $\Delta T_c$, greatly hindering the further improvement of the $T_c$ of $MgB_2$.

In this work, three types of $MgB_2$ raw materials, namely, $^aMgB_2$, $^bMgB_2$, and $^cMgB_2$, were prepared with particle sizes decreasing in order. Two types of inhomogeneous phases, namely, $Y_2O_3:Eu^{3+}$ and $Y_2O_3:Eu^{3+}/Ag$, were also prepared based on our previous preparation method [47,48]. Two other types of non-EL dopants, namely, $Y_2O_3$ and $Y_2O_3:Sm^{3+}$, were also prepared for comparison. These four types of dopants were incorporated into $MgB_2$, and the change of $T_c$ was studied. The results show that the $T_c$ of $MgB_2$ doped with non-EL $Y_2O_3$ and $Y_2O_3:Sm^{3+}$ is lower than that of pure $MgB_2$ ($\Delta T_c < 0$). By contrast, EL inhomogeneous phases $Y_2O_3:Eu^{3+}$ and $Y_2O_3:Eu^{3+}/Ag$ increase the $T_c$ ($\Delta T_c > 0$), and the optimal doping concentration of the inhomogeneous phases increased from 0.5% to 1.2% with the decrease of $MgB_2$'s particle size. The optimal doping concentrations for $^aMgB_2$, $^bMgB_2$, and $^cMgB_2$ are 0.5%, 0.8%, and 1.2%, respectively. The corresponding $\Delta T_{cs}$ are 0.4, 0.9, and 1.2 K, which exhibit significant improvements compared with the $\Delta T_{cs}$ (0.2–0.4 K) in previous work [41–45].

## 2 Model

Figure 1a–c show the cross-sectional view of MgB$_2$ SMSCs models prepared using $^a$MgB$_2$ ($\Phi_a$ < 30 μm), $^b$MgB$_2$ ($\Phi_b$ < 15 μm), and $^c$MgB$_2$ ($\Phi_c$ < 5 μm) as raw materials. $\Phi_a$, $\Phi_b$, and $\Phi_c$ refer to the particle sizes of $^a$MgB$_2$, $^b$MgB$_2$, and $^c$MgB$_2$ powders, which will be described in detail at the experiment section. The brown hexagons represent the MgB$_2$ particles, and the gray dashed lines represent the flakes of inhomogeneous phase with the surface size of approximately 20 nm and thickness of approximately 2.5 nm [45,48]. The flakes of Y$_2$O$_3$, Y$_2$O$_3$:Sm$^{3+}$, Y$_2$O$_3$:Eu$^{3+}$, and Y$_2$O$_3$:Eu$^{3+}$/Ag mainly gather on the surfaces of the MgB$_2$ particles as shown in Figure 1d. Figure 1e–h present the schematics of Y$_2$O$_3$, Y$_2$O$_3$:Sm$^{3+}$, Y$_2$O$_3$:Eu$^{3+}$, and Y$_2$O$_3$:Eu$^{3+}$/Ag, respectively. The gray flake represents Y$_2$O$_3$. The yellow, white, and green points represent Sm, Eu, and Ag. Obviously, the introduction of these four dopants inevitably reduces the $T_c$ of MgB$_2$. This is mainly because the dopants are not superconductors, which is unfavorable for the superconductivity of MgB$_2$, like the impurity phase of MgO in MgB$_2$. For convenience, the reduction in $T_c$ caused by introducing the dopants is referred to as the impurity effect [41–46]. Non-EL dopants Y$_2$O$_3$ and Y$_2$O$_3$:Sm$^{3+}$ can only decrease $T_c$ for the introduction of the impurity effect. Unlike Y$_2$O$_3$ and Y$_2$O$_3$:Sm$^{3+}$, introducing EL Y$_2$O$_3$:Eu$^{3+}$ and Y$_2$O$_3$:Eu$^{3+}$/Ag may increase the $T_c$ of MgB$_2$. In the experiment, a four-probe method is used to measure the $T_c$. During the measurements, the applied external electric field forms local electric fields in the superconductor, which could excite the inhomogeneous phase to produce EL. The generated EL excites the electrons in turn, which is favorable to the formation of Cooper pairs and enables the increase in $T_c$. This process is collectively referred to as the EL exciting effect [41–46]. A distinct competition exists between these two effects. $T_c$ would be improved ($\Delta T_c > 0$) when EL exciting effect dominates; otherwise, introducing the inhomogeneous phase would decrease $T_c$ ($\Delta T_c < 0$). During the preparation process, the impurity effect should be reduced as extensively as possible, and the EL exciting effect should be enhanced to obtain samples with a high $T_c$. The resulting superconductor is called a SMSC, and the $T_c$ of which can be improved and adjusted by incorporating EL inhomogeneous phases [41–46].

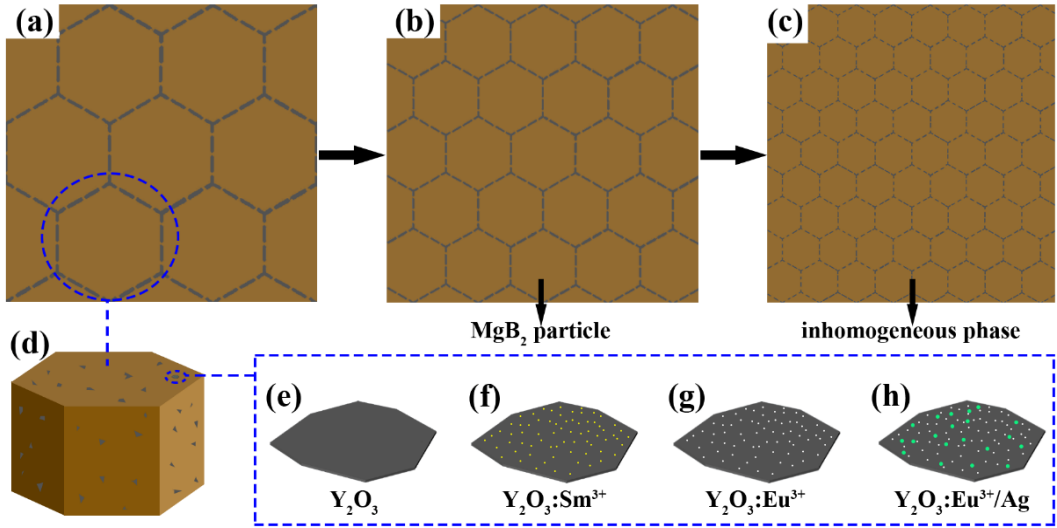

**Fig. 1** The models of MgB$_2$ SMSCs prepared using (**a**) $^a$MgB$_2$ ($\Phi_a$ < 30 μm), (**b**) $^b$MgB$_2$ ($\Phi_b$ < 15 μm), and (**c**) $^c$MgB$_2$ ($\Phi_c$ < 5 μm) as raw materials. Schematic depictions of (**d**) a particle of MgB$_2$ SMSC, (**e**) Y$_2$O$_3$, (**f**) Y$_2$O$_3$:Sm$^{3+}$, (**g**) Y$_2$O$_3$:Eu$^{3+}$, and (**h**) Y$_2$O$_3$:Eu$^{3+}$/Ag. The morphology of Y$_2$O$_3$, Y$_2$O$_3$:Sm$^{3+}$, Y$_2$O$_3$:Eu$^{3+}$, and Y$_2$O$_3$:Eu$^{3+}$/Ag is flaky with surface size of approximately 20 nm and thickness of approximately 2.5 nm [45].

However, the $ΔT_{cs}$ obtained in our previous work through the SMSC method are quite small. The low doping concentrations of inhomogeneous phases greatly hindered the further improvement of $T_c$. To further improve the $ΔT_c$ of $MgB_2$, the doping concentration of the inhomogeneous phase must be increased to enhance the EL exciting effect. However, the impurity effect inevitably increases with the increasing doping concentration, as analyzed above. The results of our previous work show that the impurity effect tends to dominate at high concentrations, which is not conducive to the $T_c$ of the sample. This phenomenon is principally caused by the agglomeration of excessive inhomogeneous phase flakes, which cannot disperse well in the sample to improve $T_c$ at concentrations exceeding the optimal value. A simple strategy to solve this problem is to reduce the particle size of $MgB_2$ as shown in Figure 1a–c. It can be seen that reducing the particle size would increase the region between the particles, thereby increasing the optimal doping concentration of the inhomogeneous phase. The inhomogeneous phase flakes can disperse well in the sample with small particle size and fully exert the EL exciting effect to further increase $ΔT_c$.

## 3 Experiment

$Y_2O_3$, $Y_2O_3:Sm^{3+}$, $Y_2O_3:Eu^{3+}$, and $Y_2O_3:Eu^{3+}/Ag$ were prepared by a hydrothermal method [47,48]. Briefly, a certain amount of $Y_2O_3$ and $Eu_2O_3$ were weighed and dissolved in HCl to make a precursor. The precursor was dissolved in benzyl alcohol and stirred on a magnetic stirrer. A certain amount of octylamine and $AgNO_3$ was added dropwise into the beaker in turn. Then the mixture was transferred to a high-pressure reaction kettle, which was then placed in a drying oven and kept at 250 °C for 24 h. Thereafter, the reaction kettle was naturally cooled to room temperature. The precipitate was washed several times with absolute ethanol to remove impurities and then separated from the solution by centrifugation, precipitation, and drying. The obtained solids were placed in a high-temperature tube furnace and heated at 800 °C for 24 h to form a white powder. After illumination, $Y_2O_3:Eu^{3+}/Ag$ was obtained. The same procedure was carried out prepare $Y_2O_3$, $Y_2O_3:Eu^{3+}$, and $Y_2O_3:Sm^{3+}$ by controlling the addition of $Eu_2O_3$ and $AgNO_3$ and replacing $Eu_2O_3$ with $Sm_2O_3$. The morphology of $Y_2O_3$, $Y_2O_3:Sm^{3+}$, $Y_2O_3:Eu^{3+}$, and $Y_2O_3:Eu^{3+}/Ag$ is flaky with surface size of approximately 20 nm and thickness of approximately 2.5 nm [45,48].

Three types of $MgB_2$ raw materials marked with $^aMgB_2$, $^bMgB_2$, and $^cMgB_2$ were prepared in this work. $Φ_a$, $Φ_b$, and $Φ_c$ refer to the particle sizes of $^aMgB_2$, $^bMgB_2$, and $^cMgB_2$ powders. A 500-mesh sieve was used to sifted $MgB_2$ powder (99%, 100 mesh, Alfa Aesar) to prepare $^aMgB_2$, indicating that $Φ_a < 30$ μm. $^bMgB_2$ was prepared by sifting $^aMgB_2$ powder through vacuum filtration with the pore size of about 15 μm, indicating that $Φ_b < 15$ μm. Meanwhile, Mg and nano boron powder sifted through vacuum filtration with the pore size of about 5 μm were applied to prepare $MgB_2$ powder by the traditional sintering process. The obtained $MgB_2$ powder was then sifted through vacuum filtration with the pore size of about 5 μm to prepare $^cMgB_2$, indicating that $Φ_c < 5$ μm. $MgB_2$-based superconductors were synthesized by an ex situ preparation process, which is described in detail in our previous work [42,45]. The doping concentrations in this work all refer to the mass percentage.

## 4 Results and discussion

Figure 2a shows the EL spectra of $Y_2O_3$, $Y_2O_3$:$Sm^{3+}$, $Y_2O_3$:$Eu^{3+}$, and $Y_2O_3$:$Eu^{3+}$/Ag, which confirm that $Y_2O_3$ and $Y_2O_3$:$Sm^{3+}$ are non-EL materials, whereas $Y_2O_3$:$Eu^{3+}$ and $Y_2O_3$:$Eu^{3+}$/Ag show a remarkable EL property. Among the four materials tested, $Y_2O_3$:$Eu^{3+}$/Ag showed the highest EL intensity because of the composite luminescence [47]. Figure 2b–d present the SEM images of the pure $MgB_2$ samples prepared using three different raw materials. Figure 2b is the SEM image of $^aMgB_2$, which shows that most of the particle exceeded 1 μm. For $^bMgB_2$, only a few of the particles exceeded 1 μm as shown in Figure 2c. Figure 2d presents the SEM image of $^cMgB_2$, which shows that most of particles are below 500 nm. The particle sizes of $^aMgB_2$, $^bMgB_2$, and $^cMgB_2$ samples decrease in order. Figure 2e reveals the XRD patterns of four samples. The black and red curves depict the XRD patterns of $^aMgB_2$ and $^aMgB_2$+0.5% $Y_2O_3$:$Eu^{3+}$/Ag, respectively. The blue and magenta curves correspond to the XRD patterns of $^bMgB_2$+0.8% $Y_2O_3$:$Eu^{3+}$/Ag and $^cMgB_2$+1.2% $Y_2O_3$:$Eu^{3+}$/Ag, respectively. The black vertical lines represent the standard XRD patterns of $MgB_2$. The main phase of all the samples was clearly $MgB_2$. The $Y_2O_3$ phase was found in the doped samples. Small amounts of the unavoidable MgO phase were also detected in all the samples [49–52]. The XRD patterns of the other samples show a similar feature.

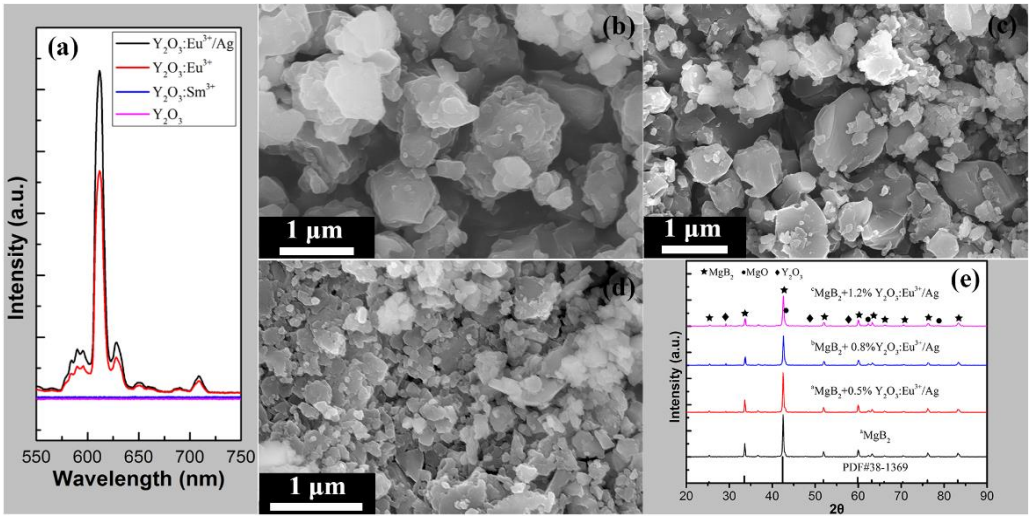

**Fig. 2** (**a**) EL intensities of $Y_2O_3$, $Y_2O_3$:$Sm^{3+}$, $Y_2O_3$:$Eu^{3+}$, and $Y_2O_3$:$Eu^{3+}$/Ag. (**b–d**) SEM images of $^aMgB_2$, $^bMgB_2$, and $^cMgB_2$, respectively. (**e**) XRD patterns of $^aMgB_2$, $^aMgB_2$+0.5% $Y_2O_3$:$Eu^{3+}$/Ag, $^bMgB_2$+0.8% $Y_2O_3$:$Eu^{3+}$/Ag, and $^cMgB_2$+1.2% $Y_2O_3$:$Eu^{3+}$/Ag.

Figure 3a illustrates the normalized resistivity-temperature (*R–T*) curves of $^aMgB_2$ doped with *x*% $Y_2O_3$ (*x* = 0, 0.2, 0.5, 0.8, 1.0, 1.2). The black curve corresponds to the $^aMgB_2$ sample, which shows that the $T_c$ of the pure sample was 37.4–38.2 K. The onset and offset critical temperatures of pure $^aMgB_2$ are 38.2 K and 37.4 K. The other curves represent $^aMgB_2$ doped with $Y_2O_3$ with concentrations of 0.2%, 0.5%, 0.8%, 1.0%, and 1.2%, indicating that the corresponding $T_{cs}$ are 37.0–37.8 K, 36.8–37.6 K, 36.5–37.3 K, 36.1–37.0 K, and 35.8–36.8 K. The results show that like conventional chemical doping, the introduction of non-EL $Y_2O_3$ decreases the $T_c$ of $MgB_2$ ($\Delta T_c < 0$) and the $T_{cs}$ of the doped samples decrease with the increase of the doping concentration as shown in the inset figure. Figure 3b shows the normalized *R–T* curves of $^aMgB_2$ doped with 0.5% *y* (*y* = 0, $Y_2O_3$, $Y_2O_3$:$Sm^{3+}$, $Y_2O_3$:$Eu^{3+}$, $Y_2O_3$:$Eu^{3+}$/Ag). The doping concentration was fixed at 0.5% base on our previous work [45]. The $T_c$ values of MgB$_2$ doped with Y$_2$O$_3$, Y$_2$O$_3$:Sm$^{3+}$, Y$_2$O$_3$:Eu$^{3+}$, and Y$_2$O$_3$:Eu$^{3+}$/Ag were 36.8–37.6 K, 36.9–37.7 K, 37.6–38.4 K, and 37.8–38.6 K. The results clearly show that non-EL Y$_2$O$_3$ and Y$_2$O$_3$:Sm$^{3+}$ decreased the $T_c$ of MgB$_2$, while EL Y$_2$O$_3$:Eu$^{3+}$ and Y$_2$O$_3$:Eu$^{3+}$/Ag increased the $T_c$ of MgB$_2$, as shown in the inset. The $T_c$ values of MgB$_2$ doped with Y$_2$O$_3$:Eu$^{3+}$ and Y$_2$O$_3$:Eu$^{3+}$/Ag increased by 0.2 and 0.4 K, respectively, compared with that of $^a$MgB$_2$. This finding is similar to those of our previous studies.

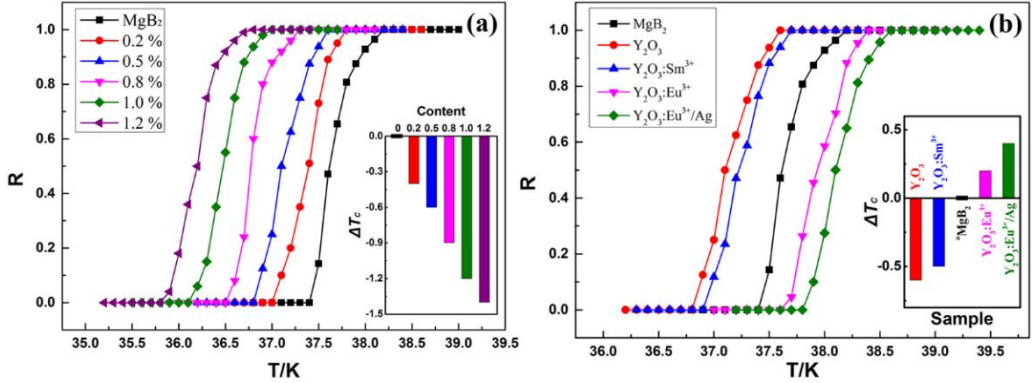

**Fig. 3** Normalized resistivity-temperature curves of $^a$MgB$_2$ doped with (**a**) $x$% Y$_2$O$_3$ ($x$ = 0, 0.2, 0.5, 0.8, 1.0, 1.2) and (**b**) 0.5% $y$ ($y$ = 0, Y$_2$O$_3$, Y$_2$O$_3$:Sm$^{3+}$, Y$_2$O$_3$:Eu$^{3+}$, Y$_2$O$_3$:Eu$^{3+}$/Ag). Insets: the values of $\Delta T_c$ ($\Delta T_c = T_c - T_{cpure}$).

Figure 4a illustrates the normalized $R$–$T$ curves of $^b$MgB$_2$ doped with $x$% Y$_2$O$_3$:Eu$^{3+}$ ($x$ = 0, 0.5, 0.6, 0.7, 0.8, 1.0). The black curve corresponds to $^b$MgB$_2$, which shows that the $T_c$ of the pure sample is 36.6–37.4 K. The other curves are the $R$–$T$ curves of $^b$MgB$_2$ doped with Y$_2$O$_3$:Eu$^{3+}$ with doping concentrations of 0.5%, 0.6%, 0.7%, 0.8%, 0.9%, and 1.0%, indicating that the corresponding $T_{cs}$ are 36.8–37.6 K, 37–37.8 K, 37.2–38.0 K, 37.4–38.2 K, 37.0–37.9 K, and 36.7–37.7 K. The $T_c$ of the doped samples first increased and then decreased with the increase of the doping concentration. The inset summarizes the evolution of $\Delta T_c$ as a function of the doping concentration. The optimal doping concentration and the corresponding $\Delta T_c$ increased to 0.8% and 0.8 K, respectively, compared with those of the samples prepared using $^a$MgB$_2$ as raw material. Figure 4b demonstrates the normalized $R$–$T$ curves of $^b$MgB$_2$ doped with 0.8% $y$ ($y$ = 0, Y$_2$O$_3$, Y$_2$O$_3$:Sm$^{3+}$, Y$_2$O$_3$:Eu$^{3+}$, Y$_2$O$_3$:Eu$^{3+}$/Ag). The $T_{cs}$ of $^b$MgB$_2$ doped with Y$_2$O$_3$, Y$_2$O$_3$:Sm$^{3+}$, Y$_2$O$_3$:Eu$^{3+}$, and Y$_2$O$_3$:Eu$^{3+}$/Ag were 35.8–36.6 K, 36.0–36.8 K, 37.4–38.2 K, and 37.5–38.3 K, respectively. Among these samples, $^b$MgB$_2$+0.8% Y$_2$O$_3$:Eu$^{3+}$/Ag obtained the highest $\Delta T_c$ (0.9 K) because of the high EL intensity, as shown in Figure 2a.

Figure 4c reveals the normalized $R$–$T$ curves of $^c$MgB$_2$ doped with $x$% Y$_2$O$_3$:Eu$^{3+}$ ($x$ = 0, 0.8, 1.0, 1.2, 1.5). Similarly, the black curve corresponds to the pure sample, indicating that the $T_c$ of $^c$MgB$_2$ is 36.0–36.8 K. The other curves correspond to $^c$MgB$_2$ doped with Y$_2$O$_3$:Eu$^{3+}$ at different concentrations of 0.8%, 1.0%, 1.2%, and 1.5%, indicating that the corresponding $T_{cs}$ are 36.2–37.0 K, 36.6–37.4 K, 37.0–37.8 K, and 36.4–37.2 K, respectively. It is same with the results in Figure 4a, that is, $T_c$ first increases and then decreases with the increase of the doping concentration, as shown in the inset figure. The optimal doping concentration is 1.2%, and the corresponding $\Delta T_c$ is 1.0 K. Figure 4d shows the normalized $R$–$T$ curves of $^c$MgB$_2$ doped with 1.2% $y$ ($y$ = 0, Y$_2$O$_3$, Y$_2$O$_3$:Sm$^{3+}$, Y$_2$O$_3$:Eu$^{3+}$, Y$_2$O$_3$:Eu$^{3+}$/Ag). The $T_c$ values of $^c$MgB$_2$ doped with Y$_2$O$_3$, Y$_2$O$_3$:Sm$^{3+}$, Y$_2$O$_3$:Eu$^{3+}$, Y$_2$O$_3$:Eu$^{3+}$/Ag are 34.7–35.7 K, 34.9–35.7 K, 37.0–37.8 K, and 37.2–38.0 K. Y$_2$O$_3$ and Y$_2$O$_3$:Sm$^{3+}$ decrease $T_c$, whereas $Y_2O_3:Eu^{3+}$ and $Y_2O_3:Eu^{3+}/Ag$ increase $T_c$. These results are consistent with those of the samples prepared using $^aMgB_2$ and $^bMgB_2$ as raw materials. The $T_c$ of $^cMgB_2+1.2\%$ $Y_2O_3:Eu^{3+}/Ag$ was enhanced by 1.2 K compared with that of the pure sample, exhibiting the highest $\Delta T_c$ among the samples.

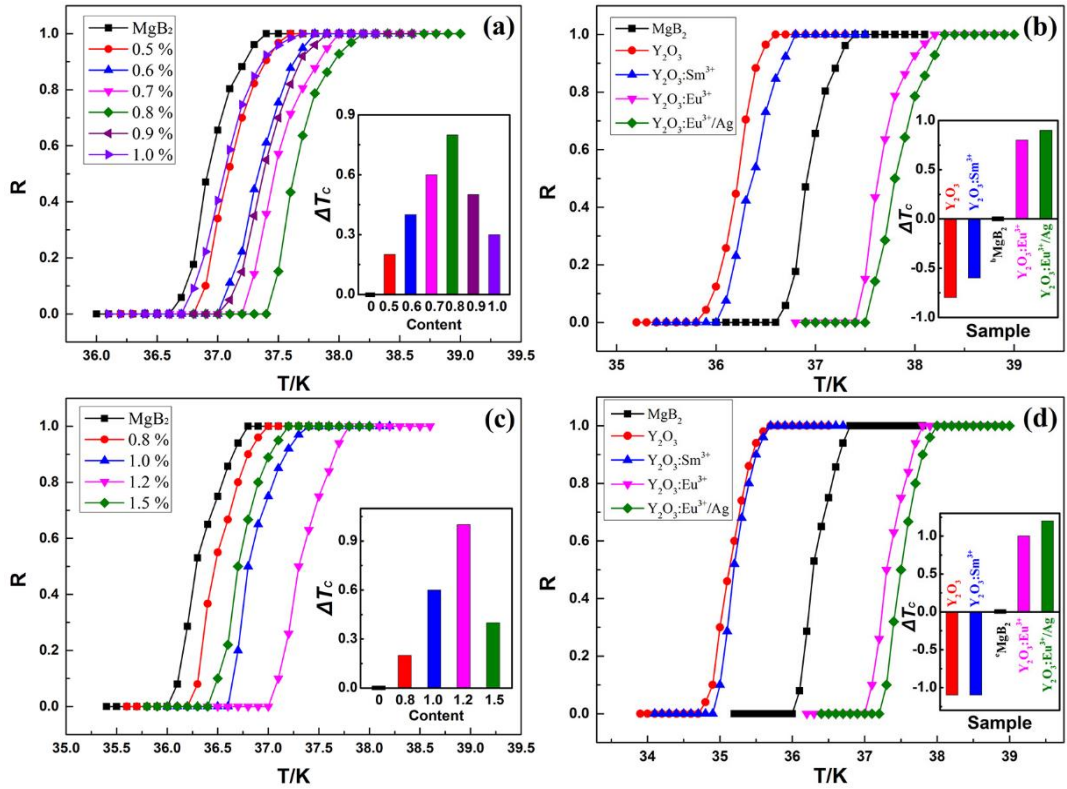

**Fig. 4** Normalized $R$–$T$ curves of $^bMgB_2$ doped with (**a**) $x\%$ $Y_2O_3:Eu^{3+}$ ($x$ = 0, 0.5, 0.6, 0.7, 0.8, 0.9, 1.0) and (**b**) 0.8% $y$ ($y$ = 0, $Y_2O_3$, $Y_2O_3:Sm^{3+}$, $Y_2O_3:Eu^{3+}$, $Y_2O_3:Eu^{3+}/Ag$). Normalized $R$–$T$ curves of $^cMgB_2$ doped with (**c**) $x\%$ $Y_2O_3:Eu^{3+}$ ($x$ = 0, 0.8, 1.0, 1.2, 1.5) and (**d**) 1.2% $y$ ($y$ = 0, $Y_2O_3$, $Y_2O_3:Sm^{3+}$, $Y_2O_3:Eu^{3+}$, $Y_2O_3:Eu^{3+}/Ag$). Insets: the values of $\Delta T_c$ ($\Delta T_c = T_c - T_{cpure}$).

Figure 5a shows the SEM image of $^aMgB_2+0.5\%$ $Y_2O_3:Eu^{3+}/Ag$. Figure 5b–e are the EDS mapping for elements Mg, Y, Eu, and Ag listed in the lower right corner of each figure. Figure 5f shows the SEM image of $^cMgB_2+1.2\%$ $Y_2O_3:Eu^{3+}/Ag$. Figure 5g–j are the EDS mapping for elements Mg, Y, Eu, and Ag. Given that the inhomogeneous phase did not react with $MgB_2$, the mapping of elements Y, Eu, and Ag can reflect the distribution of the inhomogeneous phase in the sample. It can be seen that $Y_2O_3:Eu^{3+}/Ag$ is relatively evenly distributed in $^aMgB_2$. Similarly, the inhomogeneous phase did not generate significant agglomeration in $^cMgB_2$, even though the optimal concentration was enhanced to 1.2% as the particle size decreased, as shown in Figure 5g–j. Therefore, the inhomogeneous phase was able to fully exert the EL exciting effect to further increase $\Delta T_c$ at high concentrations.

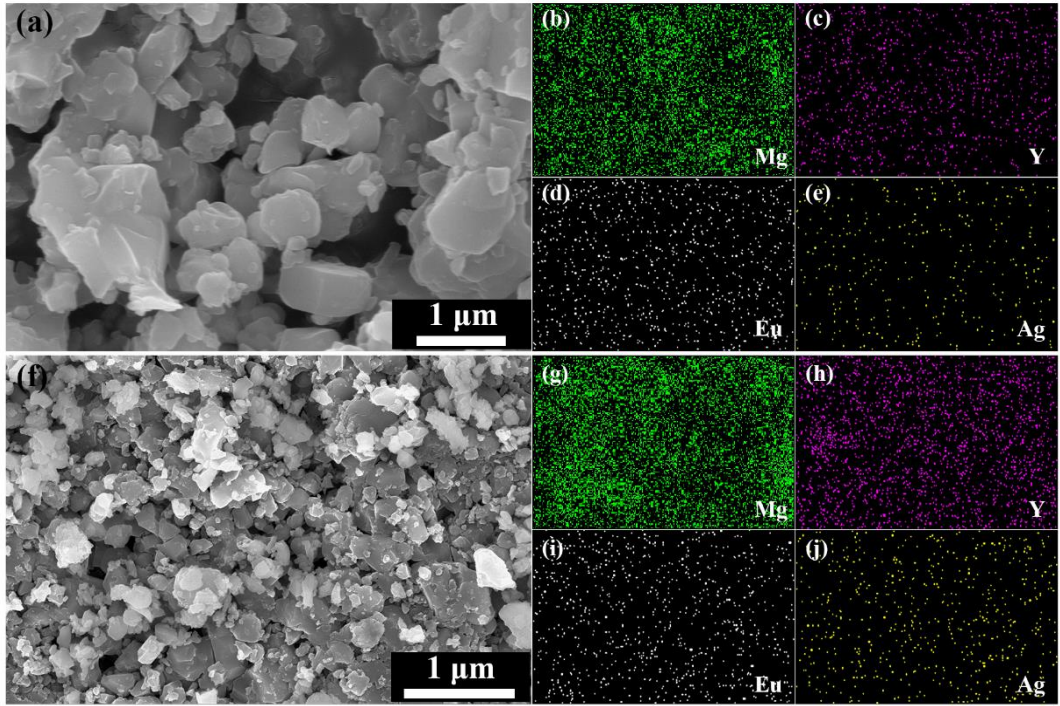

**Fig. 5** (**a**) SEM image and (**b–e**) EDS mapping of [a]MgB$_2$+0.5% Y$_2$O$_3$:Eu$^{3+}$/Ag. (**f**) SEM image and (**g–j**) EDS mapping of [c]MgB$_2$+1.2% Y$_2$O$_3$:Eu$^{3+}$/Ag.

Table 1 shows the $\Delta T_{cs}$ for [a]MgB$_2$+0.5% $x$, [b]MgB$_2$+0.8% $x$ and [c]MgB$_2$+1.2% $x$ ($x$ = Y$_2$O$_3$, Y$_2$O$_3$:Sm$^{3+}$, Y$_2$O$_3$:Eu$^{3+}$, and Y$_2$O$_3$:Eu$^{3+}$/Ag). For the [a]MgB$_2$ raw material, we prepared the MgB$_2$ SMSCs doped with 0.5% inhomogeneous phase. The results show that $\Delta T_c$ values for [a]MgB$_2$ doped with Y$_2$O$_3$:Eu$^{3+}$ and Y$_2$O$_3$:Eu$^{3+}$/Ag are 0.2 K and 0.4 K. For the [b]MgB$_2$ raw material with a smaller particle size than that of [a]MgB$_2$, the optimal doping concentration was first explored by changing the concentration of Y$_2$O$_3$:Eu$^{3+}$ from 0.5% to 1.0%. The results show that the optimal doping concentration is 0.8%. Subsequently, 0.8% Y$_2$O$_3$, Y$_2$O$_3$:Sm$^{3+}$, Y$_2$O$_3$:Eu$^{3+}$, and Y$_2$O$_3$:Eu$^{3+}$/Ag were separately doped into [b]MgB$_2$ to study the change of $T_c$. The results clearly show that Y$_2$O$_3$ and Y$_2$O$_3$:Sm$^{3+}$ reduced $T_c$, whereas Y$_2$O$_3$:Eu$^{3+}$ and Y$_2$O$_3$:Eu$^{3+}$/Ag enhanced $T_c$, and the corresponding $\Delta T_c$ values were 0.8 K and 0.9 K, respectively. Similar results were obtained in the samples prepared using [c]MgB$_2$ as the raw material. For [c]MgB$_2$, which has the smallest particle size among the three raw materials, the optimal concentration was enhanced to 1.2%. The $\Delta T_{cs}$ for [c]MgB$_2$ doped with Y$_2$O$_3$:Eu$^{3+}$ and Y$_2$O$_3$:Eu$^{3+}$/Ag were 1.0 K and 1.2 K, respectively. These results indicate that reducing the particle to increase the region between the particles can effectively enhance the optimal doping concentration, thereby enhancing the $\Delta T_c$.

**Table 1** $\Delta T_{cs}$ for [a]MgB$_2$+0.5% $x$, [b]MgB$_2$+0.8% $x$ and [c]MgB$_2$+1.2% $x$ ($x$ = Y$_2$O$_3$, Y$_2$O$_3$:Sm$^{3+}$, Y$_2$O$_3$:Eu$^{3+}$, and Y$_2$O$_3$:Eu$^{3+}$/Ag).

|  | Y$_2$O$_3$ | Y$_2$O$_3$:Sm$^{3+}$ | Y$_2$O$_3$:Eu$^{3+}$ | Y$_2$O$_3$:Eu$^{3+}$/Ag |
|---|---|---|---|---|
| [a]MgB$_2$ (0.5%) | -0.6 K | -0.5 K | 0.2 K | 0.4 K |
| [b]MgB$_2$ (0.8%) | -0.8 K | -0.6 K | 0.8 K | 0.9 K |
| [c]MgB$_2$ (1.2%) | -1.1 K | -1.1 K | 1.0 K | 1.2 K |

In this work, the $\Delta T_c$ is improved by increasing the optimal doping concentration of inhomogeneous phases through reducing the particle size, however, the $T_c$ values of MgB$_2$ SMSCs are relatively low due to the low $T_c$ of the pure MgB$_2$ sample. As the particle size decreases, the grain boundaries in the sample increase and the connectivity decreases, which are disadvantages to the superconductivity [53–55]. One possible solution is to incorporate the inhomogeneous phase into the interior of the particles to overcome the disadvantages caused by the increasing grain boundaries with the doping concentration increasing.

## 5 Conclusions

Although the effectiveness of improving the $T_c$ of superconducting materials through the SMSC method by doping with EL inhomogeneous phases has been proven in previous works, the $\Delta T_{cs}$ obtained are quite small. To further increase $\Delta T_c$, three types of MgB$_2$ raw materials, namely, $^a$MgB$_2$, $^b$MgB$_2$, and $^c$MgB$_2$, were prepared with particle sizes decreasing in order. EL inhomogeneous phases were incorporated into these three raw materials with different concentrations to study the change of $\Delta T_c$. The results show that the optimal doping concentrations for $^a$MgB$_2$, $^b$MgB$_2$, and $^c$MgB$_2$ are 0.5%, 0.8%, and 1.2%, respectively. The corresponding $\Delta T_{cs}$ are 0.4, 0.9, and 1.2 K, respectively. Meanwhile, increasing the EL intensity of the inhomogeneous phase can be considered to further increase $\Delta T_c$. This work not only proves the effectiveness of the SMSC method in improving $T_c$ but also provides an alternative approach to improving the $T_c$ of superconducting materials.

**Acknowledgement**
This work was supported by the National Natural Science Foundation of China for Distinguished Young Scholar under Grant No.50025207.

**Reference**
1. J. Bardeen, L. N. Cooper, J. R. Schrieffer, Theory of Superconductivity. Phys. Rev. **108**, 1175–1204 (1957)
2. W. L. McMillan, Transition Temperature of Strong-Coupled Superconductors. Phys. Rev. **167**, 331–344 (1968)
3. J. G. Bednorz, K. A. Müller, Possible high $T_c$ superconductivity in the Ba-La-Cu-O system. Z. Phys. B - Condensed Matter **64**, 189–193 (1986)
4. C. W. Chu, P. H. Hor, R. L. Meng, L. Gao, Z. J. Huang, Y. Q. Wang, Evidence for superconductivity above 40 K in the La-Ba-Cu-O compound system. Phys. Rev. Lett. **58**, 405–407 (1987)
5. S. N. Putilin, E. V. Antipov, O. Chmaissem, M. Marezio, Superconductivity at 94 K in HgBa$_2$CuO$_{4+\delta}$. Nature **362**, 226–228 (1993)
6. N. Loudhaief, M. Ben Salem, Structural, optical and electrical studies of Bi$_2$S$_3$ nanoparticles and their impact on the superconducting properties of (Bi,Pb)$_2$Sr$_2$Ca$_2$Cu$_3$O$_\delta$ ceramics. Appl. Phys. A **127**, (2021)
7. X. H. Chen, T. Wu, G. Wu, R. H. Liu, H. Chen, D. F. Fang, Superconductivity at 43 K in SmFeAsO$_{1-x}$F$_x$. Nature **453**, 761–762 (2008)

8. Y. Kamihara, T. Watanabe, M. Hirano, H. Hosono, Iron-Based Layered Superconductor La[O$_{1-x}$F$_X$]FeAs (x = 0.05-0.12) with T$_c$ = 26 K. J. Am. Chem. Soc. **130**, 3296–3297 (2008)
9. Z. A. Ren, W. Lu, J. Yang, W. Yi, X. L. Shen, Z. C. Li, G. C. Che, X. L. Dong, L. L. Sun, F. Zhou, Z. X. Zhao, Superconductivity at 55 K in Iron-Based F-Doped Layered Quaternary Compound Sm[O$_{1-x}$F$_x$] FeAs. Chinese Phys. Lett. **25**, 2215–2216 (2008)
10. A. Tsukada, K. E. Luna, R. H. Hammond, M. R. Beasley, J. F. Zhao, S. H. Risbud, Pulsed laser deposition conditions and superconductivity of FeSe thin films. Appl. Phys. A **104**, 311–318 (2011)
11. A. P. Drozdov, M. I. Eremets, I. A. Troyan, V. Ksenofontov, S. I. Shylin, Conventional superconductivity at 203 kelvin at high pressures in the sulfur hydride system. Nature **525**, 73–76 (2015)
12. A. Cantaluppi, M. Buzzi, G. Jotzu, D. Nicoletti, M. Mitrano, D. Pontiroli, M. Ricco, A. Perucchi, P. Di Pietro, A. Cavalleri, Pressure tuning of light-induced superconductivity in K$_3$C$_{60}$. Nat. Phys. **14**, 837–841 (2018)
13. A. P. Drozdov, P. P. Kong, V. S. Minkov, S. P. Besedin, M. A. Kuzovnikov, S. Mozaffari, L. Balicas, F. F. Balakirev, D. E. Graf, V. B. Prakapenka, E. Greenberg, D. A. Knyazev, M. Tkacz, M. I. Eremets, Superconductivity at 250 K in lanthanum hydride under high pressures. Nature **569**, 528–531 (2019)
14. Y. Sun, J. Lv, Y. Xie, H. Liu, Y. Ma, Route to a Superconducting Phase above Room Temperature in Electron-Doped Hydride Compounds under High Pressure. Phys. Rev. Lett. **123**, 097001 (2019)
15. E. Snider, N. Dasenbrock-Gammon, R. McBride, M. Debessai, H. Vindana, K. Vencatasamy, K. V. Lawler, A. Salamat, R. P. Dias, Room-temperature superconductivity in a carbonaceous sulfur hydride. Nature **586**, 373–377 (2020)
16. D. Fausti, R. I. Tobey, N. Dean, S. Kaiser, A. Dienst, M. C. Hoffmann, S. Pyon, T. Takayama, H. Takagi, A. Cavalleri, Light-Induced Superconductivity in a Stripe-Ordered Cuprate. Science **331**, 189–191 (2011)
17. A. Cavalleri, Photo-induced superconductivity. Contemp. Phys. **59**, 31–46 (2017)
18. J. Nagamatsu, N. Nakagawa, T. Muranaka, Y. Zenitani, J. Akimitsu, Superconductivity at 39 K in magnesium diboride. Nature **410**, 63–64 (2001)
19. K. P. Bohnen, R. Heid, B. Renker, Phonon dispersion and electron-phonon coupling in MgB$_2$ and AlB$_2$. Phys. Rev. Lett. **86**, 5771–5774 (2001)
20. C. Buzea, T. Yamashita, Review of the superconducting properties of MgB$_2$. Supercond. Sci. Technol. **14**, R115–R146 (2001)
21. T. Yildirim, O. Gulseren, J. W. Lynn, C. M. Brown, T. J. Udovic, Q. Huang, N. Rogado, K. A. Regan, M. A. Hayward, J. S. Slusky, T. He, M. K. Haas, P. Khalifah, K. Inumaru, R. J. Cava, Giant anharmonicity and nonlinear electron-phonon coupling in MgB$_2$: a combined first-principles calculation and neutron scattering study. Phys. Rev. Lett. **87**, 037001 (2001)
22. P. P. Singh, From E$_{2g}$ to other modes: effects of pressure on electron-phonon interaction in MgB$_2$. Phys. Rev. Lett. **97**, 247002 (2006)
23. K. Vinod, N. Varghese, U. Syamaprasad, Superconductivity of MgB$_2$ in the BCS framework with emphasis on extrinsic effects on critical temperature. Supercond. Sci. Technol. **20**, R31–R45 (2007)
24. A. Varilci, D. Yegen, M. Tassi, D. Stamopoulos, C. Terzioglu, Effect of annealing temperature on some physical properties of MgB$_2$ by using the Hall probe ac-susceptibility method. Physica B: Condensed Matter **404**, 4054–4059 (2009)
25. Y. G. Zhao, X. P. Zhang, P. T. Qiao, H. T. Zhang, S. L. Jia, B. S. Cao, M. H. Zhu, Z. H. Han, X. L. Wang, B. L. Gu, Effect of Li doping on structure and superconducting transition temperature of

Mg$_{1-x}$Li$_x$B$_2$. Physica C **361**, 91–94 (2001)

26. O. Ozturk, E. Asikuzun, S. Kaya, N. S. Koc, M. Erdem, The Effect of Ar Ambient Pressure and Annealing Duration on the Microstructure, Superconducting Properties and Activation Energies of MgB$_2$ Superconductors. J. Supercond. Nov. Magn. **30**, 1161–1169 (2016)
27. F. Cheng, Z. Ma, C. Liu, H. Li, M. Shahriar A. Hossain, Y. Bando, Y. Yamauchi, A. Fatehmulla, W. A. Farooq, Y. Liu, Enhancement of grain connectivity and critical current density in the ex-situ sintered MgB$_2$ superconductors by doping minor Cu. J. Alloy. Compd. **727**, 1105–1109 (2017)
28. Q. Zhao, C. Jiao, E. Zhu, Z. Zhu, Refinement of MgB$_2$ grains and the improvement of flux pinning in MgB$_2$ superconductor through nano-Ni addition. J. Alloy. Compd. **717**, 19–24 (2017)
29. J. C. Grivel, K. Rubešová, Increase of the critical current density of MgB$_2$ superconducting bulk samples by means of methylene blue dye additions. Physica C **565**, 1353506 (2019)
30. S. Y. Li, Y. M. Xiong, W. Q. Mo, R. Fan, C. H. Wang, X. G. Luo, Z. Sun, H. T. Zhang, L. Li, L. Z. Cao, X. H. Chen, Alkali metal substitution effffects in Mg$_{1-x}$A$_x$B$_2$ (A = Li and Na). Physica C **363**, 219–223 (2001)
31. J. S. Slusky, N. Rogado, K. A. Regan, M. A. Hayward, P. Khalifah, T. He, K. Inumaru, S. M. Loureiro, M. K. Haas, H. W. Zandbergen, R. J. Cava, Loss of superconductivity with the addition of Al to MgB$_2$ and a structural transition in Mg$_{1-x}$Al$_x$B$_2$. Nature **410**, 343–345 (2001)
32. S. X. Dou, S. Soltanian, J. Horvat, X. L. Wang, S. H. Zhou, M. Ionescu, H. K. Liu, Enhancement of the critical current density and flux pinning of MgB$_2$ superconductor by nanoparticle SiC doping. Appl. Phys. Lett. **81**, 3419–3421 (2002)
33. G. Z. Li, M. D. Sumption, M. A. Rindfleisch, C. J. Thong, M. J. Tomsic, E. W. Collings, Enhanced higher temperature (20–30 K) transport properties and irreversibility field in nano-Dy$_2$O$_3$ doped advanced internal Mg infiltration processed MgB$_2$ composites. Appl. Phys. Lett. **105**, 112603 (2014)
34. M. A. Susner, S. D. Bohnenstiehl, S. A. Dregia, M. D. Sumption, Y. Yang, J. J. Donovan, E. W. Collings, Homogeneous carbon doping of magnesium diboride by high-temperature, high-pressure synthesis. Appl. Phys. Lett. **104**, 162603 (2014)
35. M. Shahabuddin, N. A. Madhar, N. S. Alzayed, M. Asif, Uniform Dispersion and Exfoliation of Multi-Walled Carbon Nanotubes in CNT-MgB$_2$ Superconductor Composites Using Surfactants. Materials **12**, 3044 (2019)
36. H. Liu, X. P. Zhao, Y. Yang, Q. W. Li, J. Lv, Fabrication of Infrared Left‐Handed Metamaterials via Double Template‐Assisted Electrochemical Deposition. Adv. Mater. **20**, 2050–2054 (2008)
37. V. N. Smolyaninova, K. Zander, T. Gresock, C. Jensen, J. C. Prestigiacomo, M. S. Osofsky, I. I. Smolyaninov, Using metamaterial nanoengineering to triple the superconducting critical temperature of bulk aluminum. Sci. Rep. **5**, 15777 (2015)
38. I. I. Smolyaninov, V. N. Smolyaninova, Theoretical modeling of critical temperature increase in metamaterial superconductors. Phys. Rev. B **93**, 184510 (2016)
39. W. T. Jiang, Z. L. Xu, Z. Chen, X. P. Zhao, Introduce uniformly distributed ZnO nano-defects into BSCCO superconductors by nano-composite method. J. Funct. Mater **38**, 157–160 (2007) in Chinese, available at http://www.cnki.com.cn/Article/CJFDTOTAL-GNCL200701046.htm
40. S. H. Xu, Y. W. Zhou, X. P. Zhao, Research and Development of Inorganic Powder EL Materials. Materials Reports **21**, 162–166 (2007) in Chinese, available at http://www.cnki.com.cn/Article/CJFDTotal-CLDB2007S3048.htm
41. Z. W. Zhang, S. Tao, G. W. Chen, X. P. Zhao, Improving the Critical Temperature of MgB$_2$ Superconducting Metamaterials Induced by Electroluminescence. J. Supercond. Nov. Magn. **29**,

1159–1162 (2016)
42. S. Tao, Y. B. Li, G. W. Chen, X. P. Zhao, Critical Temperature of Smart Meta-superconducting $MgB_2$. J. Supercond. Nov. Magn. **30**, 1405–1411 (2017)
43. H. G. Chen, Y. B. Li, G. W. Chen, L. X. Xu, X. P. Zhao, The Effect of Inhomogeneous Phase on the Critical Temperature of Smart Meta-superconductor $MgB_2$. J. Supercond. Nov. Magn. **31**, 3175–3182 (2018)
44. Y. B. Li, H. G. Chen, W. C. Qi, G. W. Chen, X. P. Zhao, Inhomogeneous Phase Effect of Smart Meta-Superconducting $MgB_2$. J. Low. Temp. Phys. **191**, 217–227 (2018)
45. Y. B. Li, H. G. Chen, M. Z. Wang, L. X. Xu, X. P. Zhao, Smart meta-superconductor $MgB_2$ constructed by the dopant phase of luminescent nanocomposite. Sci. Rep. **9**, 14194 (2019)
46. H. G. Chen, Y. B. Li, M. Z. Wang, G. Y. Han, M. Shi, X. P. Zhao, Smart Metastructure Method for Increasing $T_c$ of Bi(Pb)SrCaCuO High-Temperature Superconductors. J. Supercond. Nov. Magn. **33**, 3015–3025 (2020)
47. M. Z. Wang, L. X. Xu, G. W. Chen, X. P. Zhao, Topological luminophor $Y_2O_3$:$Eu^{3+}$+Ag with high electroluminescence performance. ACS Appl. Mater. Interfaces **11**, 2328–2335 (2019)
48. L. X. Xu, M. Z. Wang, Z. X. Liu, X. P. Zhao, Nano-topological luminophor $Y_2O_3$:$Eu^{3+}$ + Ag with concurrent photoluminescence and electroluminescence. J. Mater. Sci.: Mater. Electron. **30**, 20243–20252 (2019)
49. D. Eyidi, O. Eibl, T. Wenzel, K. G. Nickel, M. Giovannini, A. Saccone, Phase analysis of superconducting polycrystalline $MgB_2$. Micron **34**, 85–96 (2003)
50. Q. Z. Shi, Y. C. Liu, Z. M. Gao, Q. Zhao, Formation of MgO whiskers on the surface of bulk $MgB_2$ superconductors during in situ sintering. J. Mater. Sci. **43**, 1438–1443 (2007)
51. Z. Q. Ma, Y. C. Liu, Q. Z. Shi, Q. Zhao, Z. M. Gao, The improved superconductive properties of $MgB_2$ bulks with minor Cu addition through reducing the MgO impurity. Physica C **468**, 2250–2253 (2008)
52. D. K. Singh, B. Tiwari, R. Jha, H. Kishan, V. P. S. Awana, Role of MgO impurity on the superconducting properties of $MgB_2$. Physica C **505**, 104–108 (2014)
53. M. Dogruer, G. Yildirim, O. Ozturk, C. Terzioglu, Analysis of Indentation Size Effect on Mechanical Properties of Cu-Diffused Bulk $MgB_2$ Superconductor Using Experimental and Different Theoretical Models. J. Supercond. Nov. Magn. **26**, 101–109 (2012)
54. S. Mizutani, A. Yamamoto, J.-i. Shimoyama, H. Ogino, K. Kishio, Self-sintering-assisted high intergranular connectivity in ball-milled ex situ $MgB_2$ bulks. Supercond. Sci. Technol. **27**, 114001 (2014)
55. O. Ozturk, E. Asikuzun, S. Kaya, Significant change in micro mechanical, structural and electrical properties of $MgB_2$ superconducting ceramics depending on argon ambient pressure and annealing duration. J. Mater. Sci.: Mater. Electron. **26**, 3840–3852 (2015)